\documentclass[conference]{IEEEtran}
\PassOptionsToPackage{ruled}{algorithm2e}
\usepackage[T1]{fontenc}
\usepackage[utf8]{inputenc}
\usepackage{color}
\usepackage{algorithm2e}
\usepackage{amsmath}
\usepackage{amsthm}
\usepackage{balance}
\usepackage{amssymb}
\usepackage{graphicx}
\usepackage[bookmarks=true,bookmarksnumbered=true,bookmarksopen=true,bookmarksopenlevel=1,
 breaklinks=false,pdfborder={0 0 0},pdfborderstyle={},backref=false,colorlinks=false]
 {hyperref}
\hypersetup{pdftitle={Your Title},
 pdfauthor={Your Name},
 pdfpagelayout=OneColumn, pdfnewwindow=true, pdfstartview=XYZ, plainpages=false}

\makeatletter
\theoremstyle{plain}
\newtheorem{thm}{\protect\theoremname}
\theoremstyle{plain}
\newtheorem{lem}[thm]{\protect\lemmaname}

\usepackage[caption=false,font=footnotesize]{subfig}
\usepackage{cite}

\usepackage{cite}
\usepackage{acronym}
\AtBeginDocument{\acrodef{6G}{sixth-generation}
\acrodef{AO}{alternating optimization}
\acrodef{AoA}{angle of arrival}
\acrodef{AoD}{angle of departure}
\acrodef{APGM}{alternating projected gradient method}
\acrodef{APM}{accelerated proximal gradient method}
\acrodef{AP}{access point}
\acrodef{ASP}{antenna separation product}
\acrodef{AWGN}{additive white Gaussian noise}
\acrodef{BC}{broadcast channel}
\acrodef{BCM}{block coordinate maximization}
\acrodef{BEP}{bit error probability}
\acrodef{BER}{bit error rate}
\acrodef{BF-MIMO}[BF\mbox{-}MIMO]{beamforming MIMO}
\acrodef{BF}{beamforming}
\acrodef{BS}{base station}
\acrodef{bpcu}{bits per channel use}
\acrodef{CP}{cyclic prefix}
\acrodef{CPU}{central processing unit}
\acrodef{CR}{communication rate}
\acrodef{CSI}{channel state information}
\acrodef{CSIR}{channel state information at RX}
\acrodef{SSK}{space shift keying}
\acrodef{CRLB}{Cram\'er-Rao lower bound}
\acrodef{CSIT}{channel state information at TX}
\acrodef{DCMC}{discrete\mbox{-}input continuous\mbox{-}output memoryless channel}
\acrodef{DFT}{discrete Fourier transform}
\acrodef{DL-TR-GSM}{dual-layered transmit-receive \acl{GSM}}
\acrodef{DLT}{dual-layered transmission}
\acrodef{DMA}{dynamic metasurface antenna}
\acrodef{DOA}{direction of arrival}
\acrodef{DoF}{degrees of freedom}
\acrodef{DNN}{deep neural network}
\acrodef{DPC}{dirty paper coding}
\acrodef{DRL}{deep reinforcement learning}
\acrodef{EE}{energy efficiency}
\acrodef{EGC}{equal gain combining}
\acrodef{EM}{electromagnetic}
\acrodef{EVD}{eigenvalue decomposition}
\acrodef{FPGA}{field programmable gate array}
\acrodef{FSPL}{free space path loss}
\acrodef{FFT}{fast Fourier transform}
\acrodef{FDE}{frequency domain equalization}
\acrodef{GRSM}{generalized \acl{RSM}}
\acrodef{GSM}{generalized \acl{SM}}
\acrodef{HMIMO}{holographic MIMO}
\acrodef{IA}{inner approximation}
\acrodef{IFFT}{invserse fast Fourier transform}
\acrodef{ICI}{inter-channel interference}
\acrodef{iid}[i.i.d.]{independent and identically distributed}
\acrodef{IMT}{International Mobile Telecommunications}
\acrodef{IQ}{in\mbox{-}phase and quadrature}
\acrodef{ISAC}{integrated sensing and communication}
\acrodef{ISI}{intersymbol interference}
\acrodef{ISI-free}[ISI\mbox{-}free]{intersymbol interference free}
\acrodef{LIS}{large intelligent surface}
\acrodef{LOS}{line\mbox{-}of\mbox{-}sight}
\acrodef{KKT}{Karush\mbox{-}Kuhn\mbox{-}Tucker} 
\acrodef{MA}{movable antenna}
\acrodef{MAC}{multiple-access channel}
\acrodef{mmWave}{millimeter-wave}
\acrodef{MI}{mutual information}
\acrodef{MIMO}{multiple\mbox{-}input multiple\mbox{-}output}
\acrodef{mMIMO}{massive MIMO}
\acrodef{MISO}{multiple\mbox{-}input single\mbox{-}output}
\acrodef{ML}{maximum likelihood}
\acrodef{MRC}{maximal ratio combining}
\acrodef{MMSE}{minimum mean square error}
\acrodef{MU-TR-GSM}{multiuser transmit-receive  \acl{GSM} }
\acrodef{NCSIT}{no channel state information at TX}
\acrodef{NLOS}{non\mbox{-}\acs{LOS}} 
\acrodef{NOMA}{non-orthogonal multiple access}
\acrodef{OFDM}{orthogonal frequency division multiplexing}
\acrodef{OFDMA}{orthogonal frequency division multiple access}
\acrodef{OMP}{orthogonal matching pursuit}
\acrodef{OTFS}{orthogonal time frequency space}
\acrodef{umMIMO}{ultra-massive MIMO}
\acrodef{PA}{power amplifier}
\acrodef{PAE}{power added efficiency}
\acrodef{PAPR}{peak\mbox{-}to\mbox{-}average power ratio}
\acrodef{PDF}{probability density function}
\acrodef{PEP}{pairwise error probability}
\acrodefplural{PEP}{pairwise error probabilities}
\acrodef{PGM}{projected gradient method}
\acrodef{PMP}{probability mass function}
\acrodef{PSM}{precoding-aided spatial modulation}
\acrodef{QSM}{quadrature spatial modulation}
\acrodef{RC}{reorganization computation}
\acrodef{RCS}{radar cross section}
\acrodef{RF}{radio frequency}
\acrodef{RHS}{right-hand side}
\acrodef{RIS}{reconfigurable intelligent surface}
\acrodef{RSM}{receive spatial modulation}
\acrodef{RX}{receiver}
\acrodef{SDR}{semi-definite relaxation}
\acrodef{SE}{spectral efficiency}
\acrodef{SEP}{symbol error probability}
\acrodef{SER}{symbol error rate}
\acrodef{SIC}{successive interference cancellation}
\acrodef{SIM}{stacked intelligent metasurface}
\acrodef{SINR}{signal-to-interference-plus-noise ratio}
\acrodef{SISO}{single-input single-output}
\acrodef{SM}{spatial modulation}
\acrodef{SMX-MIMO}[SMX\mbox{-}MIMO]{spatial multiplexing MIMO}
\acrodef{SMX}{spatial multiplexing}
\acrodef{SNR}{signal-to-noise ratio}
\acrodef{SC}{single carrier}
\acrodef{SCA}{successive convex approximation}
\acrodef{SVD}{singular value decomposition}
\acrodef{SPST}{single pole single-throw}
\acrodef{SR}{sensing rate}
\acrodef{SU}{secondary user}
\acrodef{TDE}{time domain equalization}
\acrodef{THz}{terraherz}
\acrodef{TX}{transmitter}
\acrodef{ULA}{uniform linear array}
\acrodef{URA}{uniform rectangular array}
\acrodef{VGA}{variable gain amplifier}
\acrodef{WSR}{weighted sum rate}
\acrodef{wrt}[w.r.t.]{with respect to}
\acrodef{ZF}{zero-forcing}
\acrodef{ZMCG}{zero-mean complex Gaussian}

}

\usepackage[skip=0pt]{caption} 
\usepackage[left=0.680in,right=0.635in,top=0.7in,bottom=1in]{geometry}

\makeatother

\providecommand{\lemmaname}{Lemma}
\providecommand{\theoremname}{Theorem}
\IEEEoverridecommandlockouts
\begin{document}
\title{Weighted Sum Rate Optimization for Movable Antenna Enabled Near-Field ISAC
\vspace{-0.5em}}

\author{\IEEEauthorblockN{Nemanja Stefan Perovi\'c\IEEEauthorrefmark{1}, Keshav Singh\IEEEauthorrefmark{1},
Chih-Peng Li\IEEEauthorrefmark{1}, and Mark F. Flanagan\IEEEauthorrefmark{2}} \IEEEauthorblockA{\IEEEauthorrefmark{1}Institute of Communications Engineering, National
Sun Yat-sen University, Kaohsiung 80424, Taiwan
}\IEEEauthorblockA{\IEEEauthorrefmark{2}School of Electrical and Electronic Engineering,
University College Dublin, Dublin 4, D04 V1W8, Ireland\\
Email: n.s.perovic@mail.nsysu.edu.tw, keshav.singh@mail.nsysu.edu.tw,
cpli@faculty.nsysu.edu.tw, mark.flanagan@ieee.org}
 \thanks{This work was supported by the National Science and Technology Council of Taiwan under Grants NSTC 114-2218-E-110-005, NSTC 112-2221-E-110-029-MY3 and NSTC 113-2222-E-110-008-MY3. }
\vspace{-2.5em}
}

\maketitle
\begin{abstract}
Integrated sensing and communication (ISAC) has been recognized as
one of the key technologies capable of simultaneously improving communication
and sensing services in future wireless networks. Moreover, the introduction
of recently developed \acp{MA} has the potential to further increase
the performance gains of ISAC systems. Achieving these gains can pose
a significant challenge for MA-enabled ISAC systems operating in the
near-field due to the corresponding spherical wave propagation. Motivated
by this, in this paper we maximize the \ac{WSR} for communication
users while maintaining a minimal sensing requirement in an MA-enabled
near-field ISAC system. To achieve this goal, we propose an algorithm
that optimizes the sensing receive combiner, the communication precoding
matrices, the sensing transmit beamformer and the positions of the
users' \acp{MA} in an alternating manner. Simulation results show
that using MAs in near-field ISAC systems provides a substantial performance
advantage compared to near-field ISAC systems with only fixed antennas.
Moreover, we demonstrate that the highest WSR is obtained when
larger weights are allocated to the users placed closer to the BS,
and that the sensing performance is significantly more affected by
the minimum sensing \ac{SINR} threshold compared to the communication
performance. \acresetall{}
\end{abstract}
\vspace{-0.65em}

\vspace{-0.65em}
\section{Introduction}
\vspace{-0.5em}
\bstctlcite{BSTcontrol}Future wireless communication systems will
have to support many new functionalities, among which high-precision
sensing is one of the most important as it enables various environment-aware
applications such as augmented reality and digital twins. A promising
technology for implementing this functionality is that of \ac{ISAC}
\cite{liu2022integrated}; this refers to a design paradigm in which
sensing and communication systems are integrated to efficiently utilize
the shared spectrum and hardware resources, while offering mutual
benefits \cite{perovic2025sensing}. As such, \ac{ISAC} is capable
of providing flexible trade-offs between the two functionalities across
various use~cases. Also, due to the possibility of integrating new wireless technologies, such as \acp{UAV} \cite{saikia2024hybrid}, and \ac{NOMA}  \cite{mondal2025outage}, ISAC has attracted significant attention from both the academic community and industry.

Simultaneously \acp{MA} have emerged as a promising technology to
further improve the effectiveness of ISAC. Dynamic positioning of
such antennas enables precise beamforming design, avoiding undesirable
side lobes and reducing interference, which enhances data rate/reliability
and sensing accuracy. Motivated by this, a significant number of papers
have studied the use of \acp{MA} in ISAC systems. In \cite{ma2024movable},
the authors studied the \ac{CRLB} for \ac{AoA} estimation as a function
of the MA positions in 1D and 2D antenna arrays, and proposed algorithms
for its optimization. The maximization of the sum of the communication
rate and the sensing \ac{MI} in a bistatic ISAC system where the
transmitter is equipped with a MA array was considered in \cite{lyu2025movable}.
In \cite{yang2025robust}, the authors proposed the design and optimization
of a \ac{RIS}-aided ISAC broadcast system with MAs at the \ac{BS}
under imperfect \ac{CSI} for both sensing and communication channels.
The practical gains achievable by using MAs with a discrete set of
possible antenna positions in an ISAC system with a dynamic radar
cross-section were analyzed in \cite{khalili2024advanced}.

In contrast to the aforementioned works that consider ISAC systems
with MAs in the far-field, ISAC systems with MAs in the near-field
remain largely unexplored. In \cite{sun2025rotatable}, the authors
proposed a sensing-centric and a communication-centric design for
near-field ISAC systems with rotatable MAs at the BS, and demonstrated
their performance advantage compared to fixed-position antennas and
non-rotatable MAs. The maximization of the weighted sum of the communication
and sensing rates in a full-duplex near-field ISAC communication system
with MAs at the BS was studied in \cite{ding2025movable}. For the
latter system, the MA positions were obtained by selecting from a
predefined set of MA positions those that provide the largest value of the objective function. Against this background, the contributions of this paper are listed
as follows:
\begin{enumerate}
\item We propose an optimization framework for a multi-user ISAC system
operating in the near field, where each user is equipped with multiple
MAs. Within this framework, we formulate an optimization problem with
the aim of maximizing the \ac{WSR} while satisfying a minimum sensing
\ac{SINR} requirement.
\item We propose an \ac{AO} based algorithm to solve the formulated problem
by decomposing it into multiple subproblems. For the sensing receive
combiner optimization, we provide a closed-form solution. The communication
precoding matrices are optimized by the \ac{SCA} method, which provides
a tight concave lower bound on the achievable rate of each user. Using
the same approach and employing \ac{SDR}, we optimize the sensing
transmit beamforming vector. Finally, the positions of the users'
MAs are obtained by the \ac{PGM} method. 
Also, we prove the
convergence of the proposed algorithm.
\item We show through simulations that the proposed algorithm achieves a
significantly improved performance compared to the benchmark scheme
where users are equipped with fixed antennas, due to a larger number
of \ac{DoF}. We demonstrate that a higher WSR is obtained when larger
rate weights are allocated to users placed closer to the BS that have
a lower free-space path loss and a larger number of DoF. Finally,
we demonstrate that the sensing performance of the considered ISAC
system is significantly more affected by the sensing SINR threshold
compared to the communication performance.
\end{enumerate}
\textcolor{black}{\emph{Notation}}\textcolor{black}{: Bold lower and
upper case letters represent vectors and matrices, respectively. $\mathbf{I}_{x}$
is the identity matrix of size \mbox{$x\times x$}. $\text{Tr}(\mathbf{X})$,
$\text{rank}(\mathbf{X})$, $||\mathbf{X}||$ and $|\mathbf{X}|$
denote the trace, rank, norm and determinant of matrix $\mathbf{X}$,
respectively. $\mathbb{E}\{\cdot\}$ stands for the expectation operator.
$\ln(\cdot)$ is the natural logarithm and $(\cdot)^{H}$ represents
Hermitian transpose. $\mathcal{CN}(\mu,\sigma^{2})$ denotes a circularly
symmetric complex Gaussian random variable with mean $\mu$ and variance
$\sigma^{2}$. $\mathfrak{R}(\cdot)$ and $\mathfrak{I}(\cdot)$ denote
the real and imaginary part of a complex number, respectively.} 

\section{System Model and Problem Formulation}
\begin{figure}[t]
\centering
\includegraphics[width=6cm, height=4.5cm]{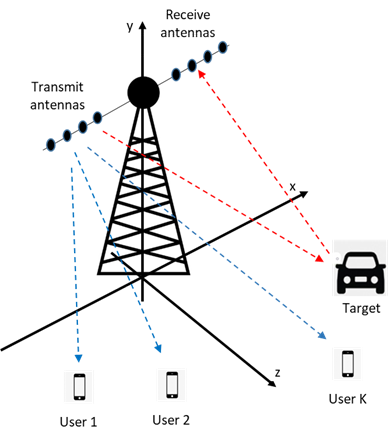}\caption{System model for the proposed MA-aided near-field ISAC system.\label{fig:sys_mod}}
\end{figure}

\vspace{-0.25em}
We consider a MA-aided near-field ISAC system, where a BS is equipped with
two fixed \acp{ULA}, as shown in Fig. \ref{fig:sys_mod}. The first BS \ac{ULA} with $N_{t}$ antennas
transmits the ISAC signal, which enables communication with $K$ users
and detection of a sensing target. The second BS ULA with $N_{r}$ antennas receives the reflected echo signal from the sensing target. The
two ULAs are separated and isolated one from another so that any mutual
coupling between antennas from different arrays is negligible. Each
user is equipped with receive MAs, such that user $k$ has $N_{k}$
MAs. Every MA of user $k$ can be moved within a 2-dimensional region
denoted by $\mathcal{C}_{k}$.

To describe the geometry of the considered system, we establish a 3D Cartesian coordinate system, where the $xz$-plane represents the
ground. The BS is placed at the origin with the ULAs in the $xy$-plane
parallel to the $x$-axis. The transmit and the receive BS ULAs have
their midpoints at $\mathbf{o}_{t}=[x_{t},y_{t},0]^{T}$ and $\mathbf{o}_{r}=[x_{r},y_{r},0]^{T}$,
and their lengths are $L_{t}$ and $L_{r}$, respectively. The coordinates
of the $m$-th ($1\le m\le N_{t}$) transmit antenna are given by
$\mathbf{t}_{m}=[x_{t,m},y_{t},0]^{T}$, where $x_{t,m}=x_{r}-L_{t}/2+(m-1)L_{t}/(N_{t}-1)$.
Similarly, the coordinates of the $n$-th ($1\le n\le N_{r}$) receive
antenna are $\mathbf{r}_{n}=[x_{r,n},y_{r},0]^{T}$, where $x_{r,n}=x_{r}-L_{r}/2+(n-1)L_{r}/(N_{r}-1)$.
Without loss of generality, we assume the MA regions of all users
are parallel to the $xy$-plane. For user $k$, the region $\mathcal{C}_{k}$
is of a square shape with the center at $\mathbf{o}_{u,k}=[x_{u,k},y_{u,k},z_{u,k}]^{T}$
and its side is denoted by $a_{k}$. The coordinates of the $N_{k}$
MAs in $\mathcal{C}_{k}$ are denoted by $\mathbf{q}_{k}=[(\mathbf{q}_{k,1})^{T},(\mathbf{q}_{k,2})^{T},\dots,(\mathbf{q}_{k,N_{k}})^{T}]^{T}$,
where $\mathbf{q}_{k,b}=[x_{k,b},y_{k,b},z_{k,b}]^{T}\in\mathcal{C}_{k}$
for $1\le b\le N_{k}$. Finally, the position of the sensing target
is specified as $\text{\ensuremath{\mathbf{s}}}=[x_{s},y_{s},z_{s}]^{T}$.

In the considered system, the transmitted ISAC signal that ensures
high-quality communication and sensing functionalities is given by
\begin{equation}
\mathbf{x}=\sum\nolimits_{k=1}^{K}\mathbf{W}_{k}\mathbf{s}_{k}+\mathbf{v}s_{0},
\end{equation}
where $\mathbf{W}_{k}\in\mathbb{C}^{N_{t}\times N_{k}}$ is the communication
precoding matrix for transmission to user $k$, and $\mathbf{s}_{k}\in\mathbb{C}^{N_{k}\times1}$
is the corresponding transmitted signal vector, which consists of
\ac{iid} symbols that are distributed according to $\mathcal{CN}(0,1)$.
Moreover, $\mathbf{v}\in\mathbb{C}^{N_{t}\times1}$ is the sensing
transmit beamformer and $s_{0}$ is the sensing signal that satisfies
$\mathbb{E}\{\left|s_{0}\right|^{2}\}=1$.

\subsection{Communication Model}

For user $k$, the received signal is given by
\begin{equation}
\mathbf{y}_{k}=\mathbf{H}_{k}\mathbf{W}_{k}\mathbf{s}_{k}+\mathbf{H}_{k}\sum\nolimits_{u=1,u\neq k}^{K}\mathbf{W}_{u}\mathbf{s}_{u}+\mathbf{H}_{k}\mathbf{v}s_{0}+\mathbf{n}_{k},\label{eq:Rec_comm_sig}
\end{equation}
where $\mathbf{H}_{k}$ is the channel matrix between the BS and user
$k$, and $\mathbf{n}_{k}\in\mathbb{C}^{N_{k}\times1}$ is the noise
vector with \ac{iid} elements distributed according to $\mathcal{CN}(0,\sigma_{k}^{2})$,
where $\sigma_{k}^{2}$ is the noise variance at user~$k$.

It is assumed that the users are located in the near field of the
BS; correspondingly, we adopt the quasi-static spherical wave channel
model for the communication channels \cite{liu2023near}. Accordingly,
the channel between the BS and user $k$ can be expressed as 
\begin{equation}
\mathbf{H}_{k}=\rho_{k}\left[\begin{array}{ccc}
e^{j\frac{2\pi}{\lambda}\|\mathbf{t}_{1}-\mathbf{q}_{k,1}\|} & \cdots & e^{j\frac{2\pi}{\lambda}\|\mathbf{t}_{N_{t}}-\mathbf{q}_{k,1}\|}\\
\vdots & \ddots & \vdots\\
e^{j\frac{2\pi}{\lambda}\|\mathbf{t}_{1}-\mathbf{q}_{k,N_{k}}\|} & \cdots & e^{j\frac{2\pi}{\lambda}\|\mathbf{t}_{N_{t}}-\mathbf{q}_{k,N_{k}}\|}
\end{array}\right],
\end{equation}
where $\lambda$ is the wavelength of operation and $\rho_{k}\in\mathbb{R}$
is the free space path-loss.

From (\ref{eq:Rec_comm_sig}), the achievable rate for user $k$ is
written as
\begin{align}
R_{k}= & \ln\Bigg|\mathbf{I}_{N_{k}}+\mathbf{H}_{k}\mathbf{W}_{k}\mathbf{W}_{k}^{H}\mathbf{H}_{k}^{H}\Biggl(\sum_{u=1,u\neq k}^{K}\mathbf{H}_{k}\mathbf{W}_{u}\mathbf{W}_{u}^{H}\mathbf{H}_{k}^{H}\nonumber \\
 & +\mathbf{H}_{k}\mathbf{v}\mathbf{v}^{H}\mathbf{H}_{k}^{H}+\sigma_{k}^{2}\mathbf{I}_{N_{k}}\Biggl)^{-1}\Bigg|.\label{eq:Rk_def}
\end{align}

Note that the communication rate in (\ref{eq:Rk_def}) is expressed
in nat/s for mathematical convenience.

\subsection{Sensing Model}

At the receive BS ULA, the receive signal is given by
\begin{equation}
\mathbf{y}_{B}=\mathbf{G}\mathbf{v}s_{0}+\mathbf{G}\sum\nolimits_{k=1}^{K}\mathbf{W}_{k}\mathbf{s}_{k}+\mathbf{z}
\end{equation}
where $\mathbf{z}\in\mathbb{C}^{N_{r}\times1}$ is the noise vector
with i.i.d. elements distributed according to $\mathcal{CN}(0,\sigma_{z}^{2})$.
For the target placed in the near-field of the BS, the sensing channel
can be expressed~as
\begin{equation}
\mathbf{G}=\rho_{s}\mathbf{f}_{r}\mathbf{f}_{t}^{H},
\end{equation}
where $\mathbf{f}_{t}=[e^{j\frac{2\pi}{\lambda}\|\mathbf{t}_{1}-\mathbf{s}\|},\dots,e^{j\frac{2\pi}{\lambda}\|\mathbf{t}_{N_{t}}-\mathbf{s}\|}]^{T}$
is the transmit near-field response vector, $\mathbf{f}_{r}=[e^{j\frac{2\pi}{\lambda}\|\mathbf{r}_{1}-\mathbf{s}\|},\dots,e^{j\frac{2\pi}{\lambda}\|\mathbf{r}_{N_{r}}-\mathbf{s}\|}]^{T}$
is the receive \mbox{near-field} response vector, and $\rho_{s}$ is
the round-trip channel coefficient.

After implementing the receive signal combining $\mathbf{u}\in\mathbb{C}^{N_{r}\times1}$,
the resulting signal is given by
\begin{equation}
y_{s}=\mathbf{u}^{H}\mathbf{y}_{B}=\mathbf{u}^{H}\mathbf{G}\mathbf{v}s_{0}+\mathbf{u}^{H}\mathbf{G}\sum\nolimits_{k=1}^{K}\mathbf{W}_{k}\mathbf{s}_{k}+\mathbf{u}^{H}\mathbf{z}.
\end{equation}
To evaluate the radar sensing capability, we use the receive SINR,
which is defined as 
\begin{equation}
\gamma_{s}=P_{s}/\text{\ensuremath{\mathbf{u}}}^{H}(\sum\nolimits_{k=1}^{K}\mathbf{G}\mathbf{W}_{k}\mathbf{W}_{k}^{H}\mathbf{G}^{H}+\sigma_{z}^{2}\mathbf{I}_{N_{k}})\text{\ensuremath{\mathbf{u}}},\label{eq:SINR_def}
\end{equation}
where $P_{s}=\text{\ensuremath{\mathbf{u}}}^{H}\mathbf{G}\mathbf{v}\mathbf{v}^{H}\mathbf{G}^{H}\text{\ensuremath{\mathbf{u}}}$
is the sensing signal power.

\subsection{Problem Formulation}

In this paper, our goal is to maximize the WSR while maintaining a
minimum SINR level $\gamma_{s}$ to guarantee the sensing performance.
Therefore, the appropriate optimization problem can be formulated
as follows: 
\begin{subequations}
\label{eq:Orig_WSR_prob}
\begin{align}
\text{\ensuremath{\underset{\mathbf{u},\{\mathbf{W}_{k}\},\mathbf{v},\{\mathbf{q}_{k}\}}{\mathrm{maximize}}}} & \;\mathrm{WSR}=\sum\nolimits_{k=1}^{K}w_{k}R_{k}\\
\text{s.t.} & \;\text{Tr}\left(\sum\nolimits_{k=1}^{K}\mathbf{W}_{k}\mathbf{W}_{k}^{H}\right)\le P_{\max}\label{eq:Pmax}\\
 & \;\gamma_{s}\ge\gamma_{0},\label{eq:SINRmin}\\
 & \;||\text{\ensuremath{\mathbf{v}}}||^{2}=1,\label{eq:TX_BF}\\
 & \;||\text{\ensuremath{\mathbf{u}}}||^{2}=1,\label{eq:RX_BF}\\
 & \;\mathbf{q}_{k}\in \mathcal{C}_{k},\label{eq:region_con}\\
 & \;||\mathbf{q}_{k,b_{1}}-\mathbf{q}_{k,b_{2}}||\ge d_{\min},1\le b_{1}\neq b_{2}\le N_{k},\text{\ensuremath{\forall}}k,\label{eq:User_space}
\end{align}
\end{subequations}
 where $w_{k}$ is the rate weight for user $k$. It should be noted
that constraint (\ref{eq:Pmax}) specifies that the total transmit
power should not exceed the available transmit power budget $P_{\max}$.
Constraint (\ref{eq:SINRmin}) specifies that the sensing SINR must
be above the SINR threshold $\gamma_{0}$. Constraints (\ref{eq:TX_BF})
and (\ref{eq:RX_BF}) enforce the unit-energy property of the sensing
transmit beamformer and the sensing receive combiner. Constraint (\ref{eq:region_con})
specifies that the MAs must remain within the region $\mathcal{C}_{k}$,
while constraint (9g) ensures the minimum inter-MA distance $d_{\min}$
for the users' MAs. This problem is difficult to solve due to the
non-convexity of the objective function, the coupling between the
optimization variables and the unit modulus constraints. To deal with
it, we propose an \ac{AO}-based algorithm which is elaborated in
the following section. \vspace{-3mm}

\section{Proposed Algorithm}

In this section, the optimization problem is decomposed into multiple
subproblems. First, we provide a closed-form expression for the sensing
receive combiner. After this, the communication precoding matrices
are optimized using a tight concave lower bound on the achievable
rate of each user, which is derived using the SCA method. The sensing
transmit beamformer is optimized by using SDR, which introduces the
covariance matrix as an optimization variable, and by applying the
SCA method to deal with the nonconvexity of the resulting problem.
The optimal positions of the users' MAs are obtained by the conventional
PGM method. Lastly, we provide the overall optimization method and
prove its convergence.

\subsection{Optimization of Sensing Receive Combiner}

Note that the optimal $\mathbf{u}$ has to maximize the sensing SINR
$\gamma_{s}$. After reformulating the numerator of $\gamma_{s}$
as
\begin{equation}
\text{\ensuremath{\mathbf{u}}}^{H}\mathbf{G}\mathbf{v}\mathbf{v}^{H}\mathbf{G}^{H}\text{\ensuremath{\mathbf{u}}}=(\rho_{s}^{2}\mathbf{f}_{t}^{H}\mathbf{v}\mathbf{v}^{H}\mathbf{f}_{t})\times(\text{\ensuremath{\mathbf{u}}}^{H}\mathbf{f}_{r}\mathbf{f}_{r}^{H}\text{\ensuremath{\mathbf{u}}}),
\end{equation}
we can observe that $\rho_{s}^{2}\mathbf{f}_{t}^{H}\mathbf{v}\mathbf{v}^{H}\mathbf{f}_{t}\ge0$
is independent of $\text{\ensuremath{\mathbf{u}}}$. Hence, the appropriate
problem can be expressed as 
\begin{subequations}
\label{eq:Opt_prob-u}
\begin{align}
\underset{\mathbf{u}}{\mathrm{maximize}} & \;\text{\ensuremath{\mathbf{u}}}^{H}\mathbf{f}_{r}\mathbf{f}_{r}^{H}\text{\ensuremath{\mathbf{u}}}/(\text{\ensuremath{\mathbf{u}}}^{H}\mathbf{D}\text{\ensuremath{\mathbf{u}}})\\
\text{s.t.} & \;(\ref{eq:RX_BF}),\nonumber 
\end{align}
\end{subequations}
 where $\mathbf{D}=\sum_{u=1}^{K}\mathbf{G}\mathbf{W}_{u}\mathbf{W}_{u}^{H}\mathbf{G}^{H}+\sigma_{z}^{2}\mathbf{I}$.
From \cite{he2023full}, the optimal $\mathbf{u}$ is given by
\begin{equation}
\text{\ensuremath{\mathbf{u}}}^{\mathrm{opt}}=\mathbf{D}^{-1}\mathbf{f}_{r}/||\mathbf{D}^{-1}\mathbf{f}_{r}||.\label{eq:u_opt}
\end{equation}

\subsection{Optimization of Communication Precoding Matrices}

We remark that the objective function in (12a) is neither convex nor
concave with respect to the precoding matrices $\{\mathbf{W}_{k}\}$.
To deal with this, we exploit the following inequality to derive a
tight concave lower bound on each user's achievable rate. For arbitrary
$p\times q$ complex matrices $\mathbf{X}$ and $\mathbf{\bar{X}}$,
and $p\times p$ complex matrices $\mathbf{Y}\succcurlyeq\mathbf{0}$
and $\mathbf{\bar{Y}}\succcurlyeq\mathbf{0}$ the following inequality
is valid \cite[Prop. 11]{zhang2023discerning}
\begin{flalign}
\ln\left|\mathbf{I}+\mathbf{X}^{H}\mathbf{Y}^{-1}\mathbf{X}\right| & \ge\ln\left|\mathbf{I}+\mathbf{\bar{X}}^{H}\bar{\mathbf{Y}}^{-1}\bar{\mathbf{X}}\right|-\text{Tr}(\mathbf{\bar{X}}^{H}\bar{\mathbf{Y}}^{-1}\bar{\mathbf{X}})\nonumber \\
 & +2\mathfrak{R}(\text{Tr}(\mathbf{\bar{X}}^{H}\bar{\mathbf{Y}}^{-1}\mathbf{X}))-\text{Tr}((\bar{\mathbf{Y}}+\mathbf{\bar{X}}\mathbf{\bar{X}}^{H})\nonumber \\
 & \times\mathbf{\bar{X}}\mathbf{\bar{X}}^{H}\bar{\mathbf{Y}}^{-1}(\mathbf{Y}+\mathbf{X}\mathbf{X}^{H})).\label{eq:Rk_LB}
\end{flalign}

Let $\{\mathbf{W}_{k}^{(n)}\}$ denote the value of $\{\mathbf{W}_{k}\}$
after $n$ iterations. After substituting $\mathbf{X}=\mathbf{H}_{k}\mathbf{W}_{k}$,
$\bar{\mathbf{X}}=\mathbf{H}_{k}\mathbf{W}_{k}^{(n)}$, $\mathbf{Y}=\sum_{u=1,u\neq k}^{K}\mathbf{H}_{k}\mathbf{W}_{u}\mathbf{W}_{u}^{H}\mathbf{H}_{k}^{H}+\mathbf{H}_{k}\mathbf{v}\mathbf{v}^{H}\mathbf{H}_{k}^{H}+\sigma_{k}^{2}\mathbf{I}_{N_{k}}$
and $\bar{\mathbf{Y}}=\sum_{u=1,u\neq k}^{K}\mathbf{H}_{k}\mathbf{W}_{u}^{(n)}(\mathbf{W}_{u}^{(n)})^{H}\mathbf{H}_{k}^{H}+\mathbf{H}_{k}\mathbf{v}\mathbf{v}^{H}\mathbf{H}_{k}^{H}+\sigma_{k}^{2}\mathbf{I}_{N_{k}}$,
and after a few simple mathematical steps, (\ref{eq:Rk_LB}) results
in
\begin{align}
R_{k} & \ge\hat{R}_{k}=\ln\left|\mathbf{I}+\mathbf{W}_{k}^{(n)H}\mathbf{H}_{k}^{H}(\mathbf{F}_{k}^{(n)})^{-1}\mathbf{H}_{k}\mathbf{W}_{k}^{(n)}\right|\nonumber \\
 & -\text{Tr}(\mathbf{W}_{k}^{(n)H}\mathbf{H}_{k}^{H}(\mathbf{F}_{k}^{(n)})^{-1}\mathbf{H}_{k}\mathbf{W}_{k}^{(n)})\nonumber \\
 & +2\mathfrak{R}(\text{Tr}(\mathbf{W}_{k}^{(n)H}\mathbf{H}_{k}^{H}(\mathbf{F}_{k}^{(n)})^{-1}\mathbf{H}_{k}\mathbf{W}_{k}))\nonumber \\
 & -\text{Tr}(\sum\nolimits_{u=1}^{K}\mathbf{W}_{u}^{H}\mathbf{H}_{k}^{H}\mathbf{A}_{k}\mathbf{H}_{k}\mathbf{W}_{u})\nonumber \\
 & -\text{Tr}(\mathbf{A}_{k}(\mathbf{H}_{k}\mathbf{v}\mathbf{v}^{H}\mathbf{H}_{k}^{H}+\sigma_{k}^{2}\mathbf{I}_{N_{k}})),
\end{align}
where $\mathbf{F}_{k}^{(n)}=\bar{\mathbf{Y}}$ and $\mathbf{A}_{k}=(\mathbf{F}_{k}^{(n)}+\mathbf{H}_{k}\mathbf{W}_{k}^{(n)}\mathbf{W}_{k}^{(n)H}\mathbf{H}_{k}^{H})^{-1}\mathbf{H}_{k}\mathbf{W}_{k}^{(n)}\mathbf{W}_{k}^{(n)H}\mathbf{H}_{k}^{H}(\mathbf{F}_{k}^{(n)})^{-1}$.

Now, it is easy to see that this lower bound is a concave function
of $\{\mathbf{W}_{k}\}$. Furthermore, the constraint (\ref{eq:SINRmin})
can be reformulated as
\begin{gather}
\!\!\!\!\mathbf{u}^{H}\mathbf{G}\left[\gamma_{0}\sum_{u=1}^{K}\mathbf{W}_{u}\mathbf{W}_{u}^{H}-\mathbf{v}\mathbf{v}^{H}\right]\mathbf{G}^{H}\mathbf{u}+\gamma_{0}\sigma_{z}^{2}\mathbf{u}^{H}\mathbf{u}\le0.\label{eq:SINR_ref}
\end{gather}

Hence, the precoding matrix optimization problem can be formulated
as 
\begin{subequations}
\label{eq:Prec_opt_prob}
\begin{align}
\underset{\{\mathbf{W}_{k}\}}{\mathrm{maximize}} & \;\widehat{\mathrm{WSR}}=\sum\nolimits_{k=1}^{K}w_{k}\hat{R}_{k}\\
\text{s.t.} & \;\text{Tr}\big(\sum\nolimits_{k=1}^{K}\mathbf{W}_{k}\mathbf{W}_{k}^{H}\big)\le P_{\max},\label{eq:Pmax-1-2}\\
 & \;\mathrm{and}\;(\ref{eq:SINR_ref}).\nonumber 
\end{align}
\end{subequations}
 This problem can be solved by a conventional optimization solver,
such as CVX.

\subsection{Optimization of Sensing Transmit Beamformer}

Similarly as in the previous subsection, we utilize the SCA method
to find the optimal $\mathbf{v}$, but we now introduce the covariance
matrix $\mathbf{V}=\mathbf{v}\mathbf{v}^{H}$. As a result, the achievable
rate of user $k$ can be written as
\begin{align}
\!\!\!\!\!\!\!R_{k} & =\ln\Bigg|\sum_{u=1}^{K}\mathbf{H}_{k}\mathbf{W}_{u}\mathbf{W}_{u}^{H}\mathbf{H}_{k}^{H}+\mathbf{H}_{k}\mathbf{V}\mathbf{H}_{k}^{H}+\sigma_{k}^{2}\mathbf{I}_{N_{k}}\Bigg|\nonumber \\ &
-\ln\Bigg|\sum_{u=1,u\neq k}^{K}\mathbf{H}_{k}\mathbf{W}_{u}\mathbf{W}_{u}^{H}\mathbf{H}_{k}^{H}+\mathbf{H}_{k}\mathbf{V}\mathbf{H}_{k}^{H}+\sigma_{k}^{2}\mathbf{I}_{N_{k}}\Bigg|.\label{eq:Rk_for_v}
\end{align}

Denoting by $\mathbf{B}_{k}(\mathbf{V})=\sum_{u=1,u\neq k}^{K}\mathbf{H}_{k}\mathbf{W}_{u}\mathbf{W}_{u}^{H}\mathbf{H}_{k}^{H}+\mathbf{H}_{k}\mathbf{V}\mathbf{H}_{k}^{H}+\sigma_{k}^{2}\mathbf{I}_{N_{k}}$
and implementing the first-order Taylor approximation to the second
term from the right-hand side, (\ref{eq:Rk_for_v}) can be lower bounded
as 
\begin{align}
R_{k} & \ge\bar{R}_{k}=\ln\left|\mathbf{B}_{k}(\mathbf{V})+\mathbf{H}_{k}\mathbf{W}_{k}\mathbf{W}_{k}^{H}\mathbf{H}_{k}^{H}\right|-\ln\left|\mathbf{B}_{k}(\mathbf{V}^{(n)})\right|\nonumber \\
 & -\text{Tr}(\mathbf{H}_{k}^{H}(\mathbf{B}_{k}(\mathbf{V}^{(n)}))^{-1}\mathbf{H}_{k}(\mathbf{V}-\mathbf{V}^{(n)})).
\end{align}

From the above, the appropriate optimization problem can be formulated
as
\begin{subequations}
\label{eq:Opt_prec_vest}
\begin{align}
\underset{\mathbf{V}}{\mathrm{maximize}} & \;\overline{\mathrm{WSR}}=\sum\nolimits_{k=1}^{K}w_{k}\bar{R}_{k}\\
\text{s.t.} & \;\text{Tr}(\mathbf{V})\le1,\label{eq:Tr_V}\\
 & \;\text{rank}(\mathbf{V})=1,\label{eq:Rank1}\\
 & \;\mathrm{and}\;(\ref{eq:SINR_ref}).\nonumber 
\end{align}
\end{subequations}
 We observe that the rank-1 constraint (\ref{eq:Rank1}) is not convex.
Also, this constraint is equivalent to the maximization of $\beta_{\max}(\mathbf{V})-\text{Tr}(\mathbf{V})$,
where $\beta_{\max}(\mathbf{V})$ is the largest eigenvalue of $\mathbf{V}$.
Exploiting the inequality $\beta_{\max}(\mathbf{V})\ge\beta_{\max}(\mathbf{V}^{(n)})+\text{Tr}(\boldsymbol{\chi}^{(n)}\boldsymbol{\chi}^{(n)H}(\mathbf{V}-\mathbf{V}^{(n)}))$,
where $\boldsymbol{\chi}^{(n)}$ is the eigenvector corresponding
to $\beta_{\max}(\mathbf{V}^{(n)})$, the previous optimization problem
can be expressed as 
\begin{subequations}
\label{eq:Opt_sense_prec_vec}
\begin{align}
\underset{\mathbf{V}}{\mathrm{maximize}} & \;\overline{\mathrm{WSR}}+\eta\mathcal{M}\\
\text{s.t.} & \;(\ref{eq:SINR_ref}),(\ref{eq:Tr_V}),\nonumber 
\end{align}
\end{subequations}
where $\mathcal{M}=\beta_{\max}(\mathbf{V}^{(n)})+\text{Tr}(\boldsymbol{\chi}^{(n)}\boldsymbol{\chi}^{(n)H}(\mathbf{V}-\mathbf{V}^{(n)}))-\text{Tr}(\mathbf{V})$,
and $\eta$ is the penalty parameter. We notice that this optimization
problem is now convex and thus it can be solved by a standard optimization
solver. \vspace{-2mm}

\subsection{Optimization of MA Positions of User $k$}

The optimization of the MA positions of user $k$ aims to maximize
the achievable rate $R_{k}$ of that user. The appropriate optimization
problem can be formulated as 
\begin{subequations}
\label{eq:Opt_qk}
\begin{align}
\underset{\mathbf{q}_{k}}{\mathrm{maximize}} & \;R_{k}\\
\text{s.t.} & \;(\ref{eq:region_con}),(\ref{eq:User_space}).\label{eq:Tr_V-1}
\end{align}
\end{subequations}
 To solve this, we use a PGM-based approach. If the MA positions of
user $k$ after $n$ iterations are denoted by $\mathbf{q}_{k}^{(n)}$,
then the positions of these antennas after $(n+1)$ iterations can
be obtained via
\begin{equation}
\mathbf{q}_{k}^{(n+1)}=P(\mathbf{q}_{k}^{(n)}+\mu_{k}^{(n)}\nabla_{\mathbf{q}_{k}}R_{k}),
\end{equation}
where $\mu_{k}^{(n)}$ is the step size. The gradient of $R_{k}$
\ac{wrt} $\mathbf{q}_{k}$ is determined by the gradients of $R_{k}$
\ac{wrt} individual MA coordinates, i.e.,
\[
\nabla_{\mathbf{q}_{k}}R_{k}=[\nabla_{x_{k,1}}R_{k},\nabla_{y_{k,1}}R_{k},0,\dots,\nabla_{x_{k,N_{k}}}R_{k},\nabla_{y_{k,N_{k}}}R_{k},0]^{T}
\]
and these gradients are provided in the following lemma.
\begin{lem}
The gradients of $R_{k}$ \ac{wrt} the x and y coordinates of the
b-th MA of user $k$ are given by (\ref{eq:dRk_x}) and (\ref{eq:dRk_y})
respectively, where $\mathbf{C}_{1,k}=\mathbf{T}_{1}\mathbf{H}_{k}^{H}\mathbf{A}_{1}^{-1},\mathbf{C}_{2,k}=\mathbf{T}_{2}\mathbf{H}_{k}^{H}\mathbf{A}_{2,k}^{-1}$\emph{,}
$\mathbf{T}_{1}=\sum_{u=1}^{K}\mathbf{W}_{u}\mathbf{W}_{u}^{H}+\mathbf{v}\mathbf{v}^{H},\mathbf{T}_{2,k}=\mathbf{T}_{1}-\mathbf{W}_{k}\mathbf{W}_{k}^{H}$,
$\mathbf{A}_{1}=\mathbf{H}_{k}\mathbf{T}_{1}\mathbf{H}_{k}^{H}+\sigma_{k}^{2}\mathbf{I}_{N_{k}}$
and $\mathbf{A}_{2,k}=\mathbf{H}_{k}\mathbf{T}_{2,k}\mathbf{H}_{k}^{H}+\sigma_{k}^{2}\mathbf{I}_{N_{k}}$.
\end{lem}
\begin{figure*}[tbh]
\begin{align}
\nabla_{x_{k,b}}R_{k} & =\frac{4\pi\rho_{k}}{\lambda}\mathfrak{I}\big(\sum\nolimits_{m=1}^{N_{t}}[\mathbf{C}_{2,k}(m,b)-\mathbf{C}_{1,k}(m,b)]\frac{x_{k,b}-x_{m}}{||\mathbf{t}_{m}-\mathbf{q}_{k,b}||}e^{j\frac{2\pi}{\lambda}||\mathbf{t}_{m}-\mathbf{q}_{k,b}||}\big)\label{eq:dRk_x}\\
\nabla_{y_{k,b}}R_{k} & =\frac{4\pi\rho_{k}}{\lambda}\mathfrak{I}\big(\sum\nolimits_{m=1}^{N_{t}}[\mathbf{C}_{2,k}(m,b)-\mathbf{C}_{1,k}(m,b)]\frac{y_{k,b}-y_{m}}{||\mathbf{t}_{m}-\mathbf{q}_{k,b}||}e^{j\frac{2\pi}{\lambda}||\mathbf{t}_{m}-\mathbf{q}_{k,b}||}\big)\label{eq:dRk_y}
\end{align}
\vspace{-1.75em}
\end{figure*}

The proof of this lemma is omitted due to space limitations. The gradient
projection is performed element-wise and independently for each MA
coordinate. For the $x$-coordinate of MA $b$, the projection onto
the moveable region is given by
\begin{equation}
P(x_{k,b})=\begin{cases}
x_{\max} & x_{\max}<x_{k,b},\\
x_{k,b} & x_{\min}\le x_{k,b}\le x_{\max},\\
x_{\min} & x_{k,b}<x_{\min},
\end{cases}
\end{equation}
where $x_{\max}=x_{u,k}+a_{k}/2$ and $x_{\min}=x_{u,k}-a_{k}/2$.
The projection for $y$ coordinates operates in a similar way. In~
each iteration, the step size $\mu_{k}^{(n)}$ is adjusted (i.e.,
decreased) using the backtracking line search until the constraint
(\ref{eq:User_space}) and the Armijo–Goldstein condition $R_{k}(\mathbf{q}_{k}^{(n+1)})-R_{k}(\mathbf{q}_{k}^{(n)})\ge\delta||\mathbf{q}_{k}^{(n+1)}-\mathbf{q}_{k}^{(n)}||^{2}$
where $\delta$ is a small positive number, are both~ satisfied.

\subsection{Overall Algorithm and Convergence Analysis}

The overall algorithm for solving the problem (\ref{eq:Orig_WSR_prob})
is summarized in Algorithm \ref{alg:Overall_alg}. Regarding the convergence
of Algorithm \ref{alg:Overall_alg}, we first observe that the optimization
$\mathbf{u}$ does not change the WSR. Moreover, $\{\mathbf{W}_{k}\}$
and $\mathbf{v}$ are optimized by using the concave lower bounds
of the objective function and implementing convex optimization. Therefore,
the appropriate objective values have to satisfy
\begin{gather}
\mathrm{WSR}(\mathbf{u}^{(n+1)},\{\mathbf{W}_{k}^{(n)}\},\mathbf{v}^{(n)},\{\mathbf{q}_{k}^{(n)}\})\nonumber \\
\le\mathrm{WSR}(\mathbf{u}^{(n+1)},\{\mathbf{W}_{k}^{(n+1)}\},\mathbf{v}^{(n)},\{\mathbf{q}_{k}^{(n)}\})\nonumber \\
\le\mathrm{WSR}(\mathbf{u}^{(n+1)},\{\mathbf{W}_{k}^{(n+1)}\},\mathbf{v}^{(n+1)},\{\mathbf{q}_{k}^{(n)}\}).
\end{gather}
The MA positions are optimized by the PGM, which always yields an
improved objective value. Based on these facts, we conclude that the
objective function is monotonically non-decreasing in each iteration
of Algorithm \ref{alg:Overall_alg}. Also, the objective function
is upper bounded due to limited communication resources. These two
facts guarantee the convergence of the objective function to a stationary
point. 
\begin{algorithm}[t]
{\footnotesize\caption{\small Proposed Algorithm for Solving (\ref{eq:Orig_WSR_prob})
\label{alg:Overall_alg}}
\DontPrintSemicolon
\LinesNumbered 

\KwIn{$\mathbf{u}^{(0)}$, $\{\mathbf{W}_{k}^{(0)}\}$, $\mathbf{v}^{(0)}$,
$\{\mathbf{q}_{k}^{(0)}\}$, $\{\mathbf{t}_{m}\}$, $\{\mathbf{r}_{n}\}$,
$\{\rho_{k}\}$, $\rho_{s}$, $\{\mu_{k}^{(0)}\}$, $d_{\min}$, $\eta$,
$\tau\in(0,1)$, $\delta>0$, $n\leftarrow0$}

\Repeat{Convergence of WSR}{

Calculate $\mathbf{u}^{(n+1)}$ using (\ref{eq:u_opt})\;

Obtain $\{\mathbf{W}_{k}^{(n+1)}\}$ by solving (\ref{eq:Prec_opt_prob})\;

Obtain $\mathbf{v}^{(n+1)}$ by solving (\ref{eq:Opt_sense_prec_vec})
\;

\For{$k=1$ $\mathrm{to}$ $K$}{

\Repeat{$R_{k}(\mathbf{q}_{k}^{(n+1)})-R_{k}(\mathbf{q}_{k}^{(n)})\ge\delta||\mathbf{q}_{k}^{(n+1)}-\mathbf{q}_{k}^{(n)}||^{2}$
$\mathrm{\mathbf{and}}$ $||\mathbf{q}_{k,b_{1}}^{(n+1)}-\mathrm{\mathbf{q}}_{k,b_{2}}^{(n+1)}||\ge d_{\min}\:(\mathrm{for}\:\mathrm{all}\:b_{1}\neq b_{2})$}{

$\mathbf{q}_{k}^{(n+1)}=P(\mathbf{q}_{k}^{(n)}+\mu^{(n)}\nabla_{\mathbf{q}_{k}}R_{k})$\;

\If{$R_{k}(\mathbf{q}_{k}^{(n+1)})-R_{k}(\mathbf{q}_{k}^{(n)})<\delta||\mathbf{q}_{k}^{(n+1)}-\mathbf{q}_{k}^{(n)}||^{2}$
$\boldsymbol{\mathbf{\mathrm{or}}}$ $||\mathrm{\mathbf{q}}_{k,b_{1}}^{(n+1)}-\mathrm{\mathbf{q}}_{k,b_{2}}^{(n+1)}||<d_{\min}\:(\mathrm{for}\:\mathrm{any}\:b_{1}\neq b_{2})$}{

$\mu_{k}^{(n)}\leftarrow\tau\mu_{k}^{(n)}$\;

}}

}

$n\leftarrow n+1$ \;

}

\KwOut{$\mathbf{u}^{\mathrm{opt}}$, $\{\mathbf{W}_{k}^{\mathrm{opt}}\}$,
$\mathbf{v}^{\mathrm{opt}}$, $\{\mathbf{q}_{k}^{\mathrm{opt}}\}$}}
\end{algorithm}

\section{Simulation Results}

In this section, we evaluate the WSR of the considered system using
the proposed algorithm by means of Monte Carlo simulations. As a benchmark,
we consider a scheme that differs from the proposed one in that users
have fixed receive antennas, and is referred to as \emph{Fix-User-Ant}.
The initial positions of the users' MAs in the proposed scheme are
randomly chosen inside the MA region $\mathcal{C}_{k}$ and are the
same as the positions of the users' antennas in \emph{Fix-User-Ant.}
The free space path loss between the BS and user $k$ modeled as $\rho_{k}=\lambda^{2}/(4\pi d_{tk})^{2}$
\cite{liu2023near}, where $d_{tk}=\|\mathbf{o}_{t}-\mathbf{o}_{k}\|$
is the distance between the BS and user $k$. The round-trip channel
coefficient for target sensing is given by $\rho_{s}=\lambda^{2}/((4\pi)^{3}R_{t}^{2}R_{r}^{2})$
\cite{dong2022sensing}, where $R_{t}=\|\mathbf{o}_{t}-\mathbf{s}\|$
is the distance between the transmit BS MA array and the target, and
$R_{r}=\|\mathbf{o}_{r}-\mathbf{s}\|$ is the distance between the
receive BS MA array and the target.

In the following simulation setup, the parameters are $f=30\,\mathrm{GHz}$
(i.e., $\lambda=1\,\mathrm{cm}$), $N_{t}=8$, $N_{r}=8$, $K=2$,
$N_{k}=2$ (for $k=1,2$), $w_{k}=1/2$ (for $k=1,2$), $P_{\max}=1\,\mathrm{W}$, $d_{\min}=\lambda/2=0.5\,\mathrm{cm}$,
$\gamma_{0}=0.01$, $\eta=1$ and $\sigma_{k}^{2}=\sigma_{z}^{2}=-100\,\mathrm{dB}$.
The midpoints of the transmit and the receive BS ULAs are located
at $\mathbf{o}_{t}=[-3\,\mathrm{m},10\,\mathrm{m},0]^{T}$ and $\mathbf{o}_{r}=[3\,\mathrm{m},10\,\mathrm{m},0]^{T}$,
respectively, and the lengths of both ULAs are $L_{t}=L_{r}=1\,\mathrm{m}$.
The centers of the users' MA regions are located at $\mathbf{o}_{u,1}=[-4\,\mathrm{m},1.5\,\mathrm{m},15\,\mathrm{m}]^{T}$
and $\mathbf{o}_{u,2}=[2\,\mathrm{m},1.5\,\mathrm{m},20\,\mathrm{m}]^{T}$,
and the side length of both square regions is $a_{k}=20\,\mathrm{cm}$.
The initial positions of the users' antennas are randomly selected
inside the specified MA regions. The point target is located at $\mathbf{s}=[10\,\mathrm{m},1.5\,\mathrm{m},10\,\mathrm{m}]^{T}$.
The CVX tool is used to solve (\ref{eq:Prec_opt_prob}) and (\ref{eq:Opt_sense_prec_vec}).
In the line search procedure for the PGM, all step sizes are initially
set to $1$, $\delta=10^{-2}$ and $\tau=1/2$. All results are averaged
over 50 independent channel realizations.

\begin{figure}[t]
\centering{}\includegraphics[width=7cm,height=4.5cm]{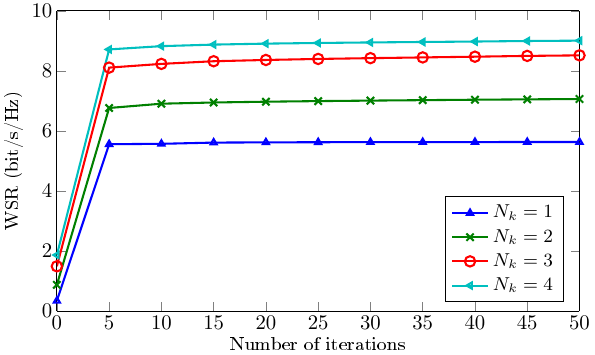}\caption{Convergence of the proposed algorithm. \label{fig:WSR-conv}}
\end{figure}

The convergence of the proposed algorithm for different numbers of
MAs at the users is shown in Fig. \ref{fig:WSR-conv}. In general,
we can see that the proposed algorithm requires less than 20 iterations
to converge regardless of the number of user MAs. The WSR increases
with $N_{k}$ since adding additional MAs improves the number of DoF,
and this effect is usually more pronounced in near-field communications.
However, the~rate~of this increase reduces when $N_{k}$ changes from
3 to 4~due~to the logarithmic nature of the user's
achievable rate expression.
\begin{figure}[t]
\centering{}\includegraphics[width=7cm,height=4.5cm]{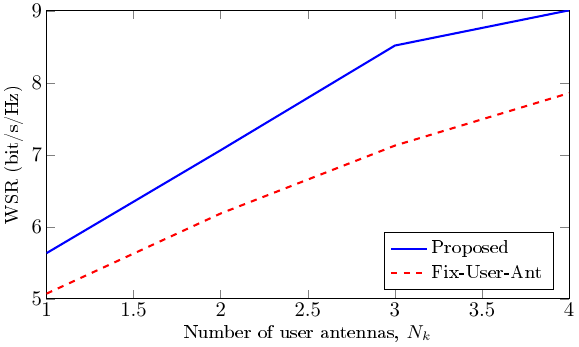}\caption{WSR versus the number of user antennas. \label{fig:WSR-vs-Nk}}
\end{figure}

In Fig. \ref{fig:WSR-vs-Nk}, we present the WSR for different numbers
of MAs at each user. The proposed scheme achieves a larger WSR compared
to the \emph{Fix-User-Ant} scheme, and this performance advantage
is gradually increased with $N_{k}$. The main reason is that changing
the MAs positions can actually reduce any correlation between the
channels of the receive MAs at each user, which in turn improves the
number of DoF and each user's achievable rate. However, for larger
WSR values, the logarithmic nature of the user's achievable rate expression
can reduce any user's achievable rate improvement caused by the MAs
movements, and therefore the WSR advantage of the proposed scheme
decreases when $N_{k}$ changes from 3 to 4.

\begin{figure}[t]
\centering{}\includegraphics[width=7cm,height=4.5cm]{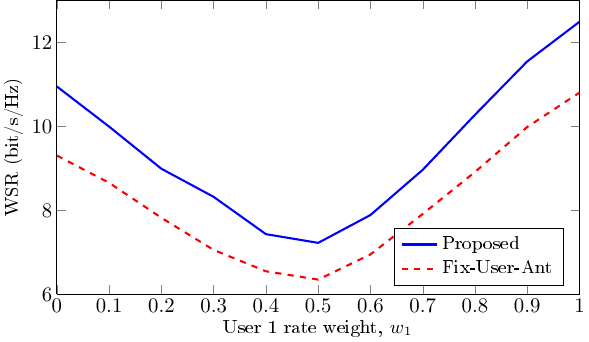}\caption{WSR versus the users rate weights. \label{fig:WSR-vs-weight}}
\end{figure}

The WSR for different values of the users' rate weights is presented
in Fig. \ref{fig:WSR-vs-weight}, where we have imposed the condition
$w_{1}+w_{2}=1$. It can be observed that both schemes achieve the
largest WSR when the data transmission is entirely allocated to user
1 (i.e., $w_{1}=1$). Even for $w_{1}>0.85$, when most of data transmission
is allocated to user 1, the WSR is greater than in the case when the
data transmission is entirely allocated to user 2 (i.e., $w_{1}=0$).
This is because user 1 is located closer to the transmit BS MAs than
user 2, which results in a lower free-space path loss and a larger
achievable rate of user 1. Moreover, a user located closer to the
BS generally has a larger number of DoF, which also contributes to
the increased achievable rate of that user. On the other hand, the
lowest WSR occurs when $w_{1}$ lies around 0.5, when the weighted
rate contributions of both users to the WSR are approximately equal.
\begin{figure}[t]
\centering{}\includegraphics[width=7cm,height=4.5cm]{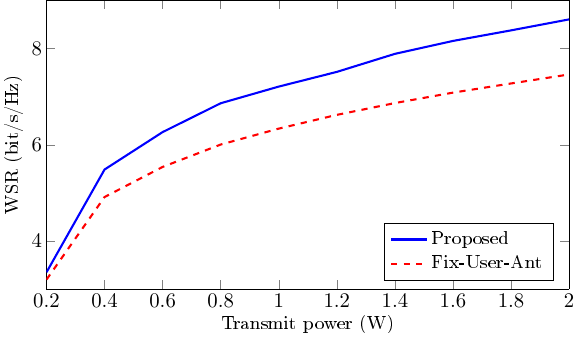}\caption{WSR versus the maximum transmit power. \label{fig:WSR-vs-power}}
\end{figure}

Next, we study how the WSR varies with the maximum transmit power
(i.e., $P_{\max}$), as shown in Fig. \ref{fig:WSR-vs-power}. The
WSR curves for both schemes exhibit an approximately logarithmic shape
due to the logarithmic increase in the users' achievable rates. As
expected, the proposed scheme with MAs achieves a higher WSR compared
to the \emph{Fix-User-Ant} scheme. The WSR performance advantage of
the proposed scheme is gradually increased with the transmit power,
which implies that the use of MAs is particularly beneficial in the
high transmit power~range.

\begin{figure}[t]
\centering{}\includegraphics[width=7cm,height=4.5cm]{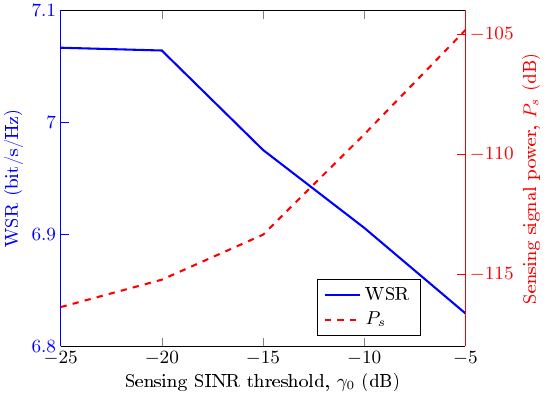}\caption{WSR and sensing signal power versus the sensing SINR threshold.\label{fig:WSR-and-sensing}}
\end{figure}

In Fig. \ref{fig:WSR-and-sensing}, we show the WSR and the sensing
signal power $P_{s}$ versus the sensing SINR threshold $\gamma_{0}$.
In general, we notice that the considered sensing metrics is far more
sensitive to the change of the sensing SINR threshold compared the
considered communication metric. More precisely, the sensing signal
power shows a significant increase for more than $10\,\mathrm{dB}$
with the sensing SINR threshold. On the other hand, the WSR remains
almost unchanged when the sensing SINR threshold is below $-20\,\mathrm{dB}$
and has a relatively modest decrease for larger SINR threshold values.

\vspace{-0.65em}
\section{Conclusion}
\vspace{-0.25em}
In this paper, we have studied the WSR maximization in a near-field ISAC system, where each user is equipped with MAs. To solve it, we developed an AO-based algorithm that alternately optimizes the sensing receive combiner, the communication precoding matrices, the sensing transmit beamformer and the positions of the users' MAs. The convergence of the proposed algorithm was also shown. Simulation results verified the effectiveness of the proposed scheme, which achieves around $11\thinspace\%$,
$14\thinspace\%$, $20\thinspace\%$ and $15\thinspace\%$ larger WSR compared to a fixed-antenna benchmark scheme in a near-field ISAC system with 1, 2, 3 and 4 antennas at each user, respectively. Furthermore, it can be observed that higher WSRs are obtained when larger weight rates are allocated to the users that are placed closer to the BS, and that the sensing SINR threshold significantly more influences the sensing than the communication features of ISAC.

\balance
\bibliographystyle{IEEEtran}
\bgroup\inputencoding{latin9}\bibliography{IEEEabrv,IEEEexample,references}
\egroup

\end{document}